\documentstyle[twoside,fleqn,espcrc2]{article}
\title{
\vspace{-2cm}
\begin{flushright}
KANAZAWA 99-18\ \ \ \ \ \ \\
August 1999\ \ \ \ \ \ \  
\end{flushright}
\vspace*{-.5cm}
Recent topics of infrared effective lattice QCD
\thanks{The authors of each part are 1. K.Yamagishi, S.Kitahara and T.Suzuki, 
2. T.Suzuki, S.Ito, T.Tsunemi, S.Fujimoto and S.Kato, 3.  F.Shoji,
 T. Suzuki, H. Kodama and A. Nakamura}}
\author{Tsuneo Suzuki
\address{Department of Physics, Kanazawa University, Kanazawa 920-1192, 
Japan}, Shun-ichi Kitahara
\address{Jumonji University, Niiza,Saitama 352-8510, Japan},
Fumiyoshi Shoji\address{Research Institute for Information Science and 
Education, Hiroshima University, Higashi-Hiroshima 739-8521, Japan},
Atsushi Nakamura \hspace{-2mm}\addtocounter{address}{-1}
\addressmark,
 Kentarou Yamagishi \hspace{-2mm}\addtocounter{address}{-3} 
\addressmark,
Shoichi Ito \hspace{-2mm}\addtocounter{address}{-1} 
\addressmark,
 Tomohiro Tsunemi \hspace{-2mm}\addtocounter{address}{-1} 
\addressmark, 
Shouji Fujimoto \hspace{-2mm}\addtocounter{address}{-1} 
\addressmark,
 Seikou Kato \hspace{-2mm}\addtocounter{address}{-1} 
\addressmark and
Hiroaki Kodama \hspace{-2mm}\addtocounter{address}{-1} 
\addressmark}

\begin{document}

\begin{abstract}
Three topics concerning infrared effective lattice QCD are discussed.
(1)Perfect lattice action of infrared $SU(3)$ QCD and perfect 
operators for the static potential are analytically given when we assume 
two-point monopole interactions alone. The assumption seems to be justified 
from numerical analyses of pure $SU(3)$ QCD in maximally abelian gauge. 
(2)Gauge invariance of monopole dominance can be
 proved theoretically if the gauge invariance of abelian dominance is
 proved. 
The gauge invariance of monopole condensation leads us to confinement of 
abelian neutral but color octet states after abelian projection. 
(3)A stochastic gauge fixing method is developed to 
study the gauge dependence of the Abelian projection, which interpolates
 between the maximally abelian (MA) gauge and no gauge fixing.
Abelian dominance for the heavy quark potential 
holds even in the gauge which is far from Maximally Abelian one.
\end{abstract}

\maketitle

\input epsf

\vspace{-.5cm}

\section{SU3 infrared effective monopole action
}

\begin{figure}[tb]
\epsfxsize=0.4\textwidth
 \begin{center}
\vspace*{-4cm}
  \epsfbox[0 133 595 729]{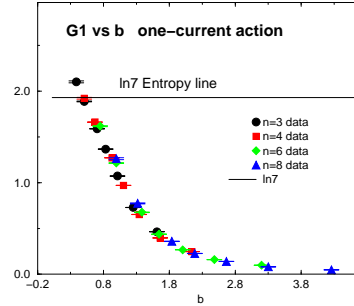}
 \end{center}
\vspace{-1.cm}
\caption{
The self coupling in the one-current case.
}
\label{G1}
\vspace{-.5cm}
\end{figure}

Abelian dominance and monopole dominance in MA gauge suggest the
existence of an effective monopole action also in $SU(3)$ QCD. 
There are two independent (three with one constraint 
$\sum_{i=1}^{3}k^i_{\mu}(s)=0$) currents in the case of $SU(3)$.
Applying the same inverse Monte-Carlo method developed in the case of
$SU(2)$ QCD\cite{shiba95,nakamura98},  
we have derived the infrared effective monopole action starting from 
the vacuum ensemble \{$k_{\mu}^i(s)$\} defined from the thermalized abelian 
link fields after the MA abelian projection\cite{degrand}.
Effective monopole actions can be derived similarly for the blocked 
monopole currents $K_{\mu}^\alpha(s)=\sum_{i,j,m=0}^{n-1} k_{\mu}^\alpha
(ns+(n-1)\hat{\mu}+i\hat{\nu}+j\hat{\rho}+m\hat{\sigma})$\cite{ivanenko}.
This corresponds to the block-spin transformation of the monopole
currents on the dual lattice. The form of the action is restricted to 
27 two-point interactions up to $3na(\beta)$ distance 
between monopole currents and the most leading 4 and 6 point
interactions. The lattice size is $48^4$ for $\beta=5.6\sim 6.4$.
The results are as follow:
\begin{enumerate}
\vspace*{-.3cm}
\item
The monopole actions are fixed clearly when we pay attention to 
two independent monopole currents: $S(k)=\sum_i G_i(S_i(k^1)+S_i(k^2)
+S_i(-k^1-k^2)$ as well as  one kind of monopole currents:
$S(k)=\sum_{i}G_i S_i(k)$.
\vspace*{-.3cm}
\item
The scaling $S(n,a(\beta))=S(b=na(\beta))$ for fixed $b=na(\beta)$ 
 looks good both for one-current and two-current cases. (Fig.\ref{G1})
\vspace*{-.3cm}
\item
Monopole condensation occurs in the large $b$ region.
Entropy dominates over the energy: ln7$> G_1$ in
the  one-current case. (Fig.\ref{G1})
\vspace*{-.3cm}
\item
The quadratic monopole actions are dominant for large $b$ in both cases.
\end{enumerate}

\vspace{-.2cm}
In the infrared regions, quadratic monopole interactions are 
dominant. When we restrict ourselves to quadratic monopole interactions 
alone, the block-spin transformation can be done analytically 
as done in $SU(2)$ QCD\cite{suzuki99}. 
We can obtain also the perfect operator to evaluate the static
quark-antiquark potential $V(Ib,Jb,Kb)$ on the coarse $b$ lattice. 

In the framework of the string model, the quantum effects are 
seen to be small in the infrared region and 
the static potential is evaluated by the classical 
contributions alone:
\begin{eqnarray*}
\langle W(C)\rangle_{cl}&=&\exp\{-\frac23 \pi^2
a^8\sum_{s,s',\mu,\nu}N_\mu(as)\\
&\times& A_{\mu\nu}(as-as')N_\nu(as')\},
\end{eqnarray*}
where $N_\mu(s)=\sum\Delta^{-1}(s-s')\frac12\epsilon_{\mu\alpha\beta\gamma}
\partial S_{\beta\gamma}(s'+\hat\mu)$ and $S_{\beta\gamma}(s)$ is the 
source term corresponding to the Wilson loop. 
We get the rotational invariance $V(Ib,Ib,0)/V(Ib,0,0)=\sqrt{2}$ and 
$\sigma=\pi\kappa\ln(m_1/m_2)/3$, where 
$m_1$ and $m_2$ are  dimensional constants parametrizing 
the monopole action on the fine $a$ lattice.
$\sigma$ is equivalent to the result from DGL theory\cite{suzuki89}.
\begin{figure}[htb]
\vspace{-1cm}
\epsfxsize=0.4\textwidth
 \begin{center}
 \leavevmode
  \epsfbox[20 133 615 729]{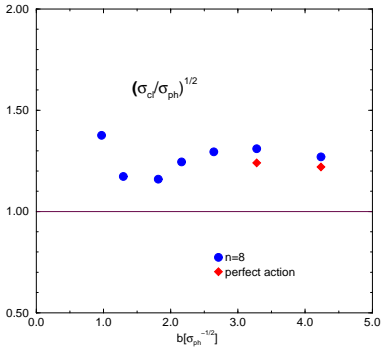}
 \end{center}
\vspace{-3.cm}
\caption{
The string tension evaluated from the $SU(3)$ effective action.
}
\label{string}
\vspace{-.5cm}
\end{figure}

\vspace{-.5cm}

\section{
Gauge independence of abelian and monopole dominances
}

Abelian and monopole dominances are shown numerically in MA gauge. 
However,
monopole condensation after abelian projection in some gauges 
{\it can not explain color confinement}.
 Abelian neutral but color non-singlet states are
not confined. 
{\bf Gauge independence of monopole condensation is essential for color 
confinement.}
Here we prove the gauge independence of the monopole dominance 
if abelian dominance is gauge independent.

First we note that abelian Wilson loops  
without any gauge fixing are known to give the same string
tension \cite{yotsu90,faber98} as that from non-abelian Wilson loops 
in the strong coupling expansion. Recently, 
Ogilivie\cite{ogilvie} proved the same fact for any $\beta$ using 
the character expansion\cite{faber98}.
Abelian dominance comes from the Elitzur's theorem\cite{elitzur75} 
and then it must be gauge independent.

Abelian dominance leads us to abelian effective action.
We express the abelian action in terms of the Villain form
$Z=\int_{-\pi}^{\pi} D \theta 
\sum_{n\in {\bf Z}}{\rm e}^{-F[d\theta + 2\pi n]}.$
The general Villain action can be expressed as follows:
\begin{eqnarray*}
&&\hspace*{-.5cm}
{\rm e}^{-(d\theta + 2\pi n,D(d\theta + 2\pi n))
          - F^{'}[d\theta + 2\pi n]} \\
&&\hspace*{-.5cm}=e^{-F^{'}[-i\delta/\delta B]}
{\rm e}^{-(d\theta + 2\pi n ,D(d\theta + 2\pi n))
          + i(B,d\theta + 2\pi n)}\bigg|_{B=0},
\end{eqnarray*}
where $[D,d]=[D,\delta]=0$ are satisfied in the large $\beta$ scaling region.

The abelian Wilson loop ${\rm e}^{i(\theta ,J)}$ is estimated 
with this action, where $J$ is the electric current.
When we use the BKT transformation\cite{BKT}, we get
an action in terms of monopole currents:
\begin{eqnarray*}
&&\hspace*{-.2cm}Z(J) = e^{-F^{'}[-i\delta/\delta B]}\sum_{k\in {\bf Z}, dk=0}
{\rm e}^{-\frac{1}{4}( \delta B , (\Delta D )^{-1}
  \delta B )} \\
&&
{\rm e}^{\frac{1}{2}( i B , 4\pi \delta \Delta^{-1}k
+ i  (\Delta D )^{-1} dJ )} 
{\rm e}^{-4\pi^{2} ( k , \Delta^{-1}D  k )}\\
&&
{\rm e}^{2\pi i ( \delta \Delta^{-1}k , S )}
{\rm e}^{- \frac{1}{4}( J , ( \Delta D )^{-1} J )}
\bigg|_{B=0} 
\end{eqnarray*}
1)Electric-electric current $J-J$ interactions (with no monopole $k$) 
come from the exchange
of regular photons and have no line singularity leading to a linear 
potential. Hence
the linear potential of abelian Wilson loops is due to 
the monopole contribution alone. 
{\bf Monopole dominance is proved 
from abelian dominance.}
\noindent
2)The linear potential comes only from  
$\exp(2\pi i (\delta \Delta^{-1}k , S))$. 
The surface independence of the 
static potential is assured due to the 4-d linking number.

\vspace{-.3cm}
\section{
A new gauge fixing method for abelian projection
}

To confirm the above results numerically, we analyze gauge dependence
of abelian projection by
Langevin equation with stochastic gauge fixing term\cite{zwanzger},
\begin{equation}
\frac{\partial}{\partial\tau}A^a_\mu(x,\tau)=
-\frac{\delta S}{\delta
A^a_\mu(x,\tau)} + \frac{1}{\alpha}D^{ab}_\mu \Delta^b 
 + \eta^a_\mu(x,\tau),
\label{SQE}
\end{equation}
where $\tau$ is fictious time,
$\eta$ is Gaussian white noise and $\Delta$ is gauge fixing term.
$\alpha=0$ ($\alpha=\infty$) corresponds to the MA gauge fixing
(no gauge fixing).

A compact lattice version of Eq.(\ref{SQE}) 
was proposed in Ref.\cite{mizu},
\[
U_\mu(x,\tau+\Delta\tau)=
\omega^{\dagger}(x,\tau)e^{if^a_\mu t^a}
U_\mu(x,\tau)\omega(x+\hat{\mu},\tau), 
\]
\vspace{-.5cm}
\[
f_\mu^a=-\frac{\partial S}{\partial
A^a_\mu}\Delta\tau+\eta^a_\mu(x,\tau)
\sqrt{\Delta\tau},\nonumber 
\]
\vspace{-.5cm}
\[
\omega(x,\tau)={\rm exp}(i\beta\Delta^a(x,\tau) 
t^a \Delta\tau/2N_c\alpha).\nonumber 
\]
We set 
$ \Delta(x,\tau)=i[\sigma_3,X(x,\tau)] $,
where $X$ is the operator to be diagonalized in MA gauge.

\begin{figure}[htb]
\vspace{-3.5cm}
\epsfxsize=0.4\textwidth
 \begin{center}
 \leavevmode
  \epsfbox[20 133 615 729]{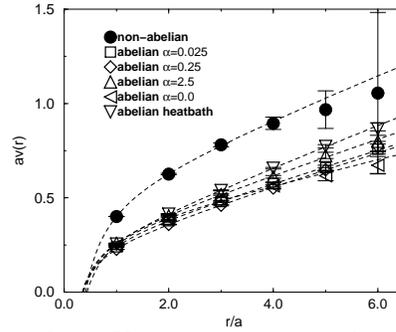}
 \end{center}
\vspace{-0.5cm}
\caption{
Heavy quark potentials from non-abelian and abelian contributions.
}
\label{pot-abel}
\vspace{-.5cm}
\end{figure}

As a first test, we perform
numerical simulations on a $12^4$ lattice with
$\beta=2.35$, $\alpha=0.025,0.25,2.5$, and  $\Delta\tau=0.001$.
We show the abelian heavy quark potentials for different $\alpha$
in Fig.\ref{pot-abel} together with that of non-abelian potential,
which are essentially same from $\alpha=0$ to 2.5,
and show the confinement linear potential behavior.

Note that the string tension obtained by the standard
gauge fixing procedure is larger than that of $\alpha=0$.
This is probably due to the Gribov copies\cite{hioki}.  Bali et al. \cite{bali}
have observed that when the number of copies decreases the string tension
becomes small. This is very consistent with our data.

\begin{figure}[htb]
\vspace{-3.5cm}
\epsfxsize=0.4\textwidth
 \begin{center}
 \leavevmode
  \epsfbox[20 133 615 729]{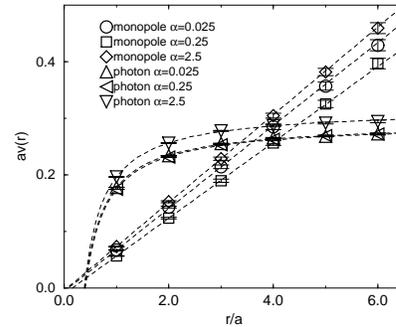}
 \end{center}
\vspace{-0.5cm}
\caption{
Heavy quark potentials from monopole and photon contributions.
}
\label{pot-mono}
\end{figure}

The heavy quark potentials from monopole and photon contributions are
plotted in Fig.\ref{pot-mono}.
The monopole part shows the confinement behavior.


T.S. acknowledges financial support from JSPS 
Grant-in Aid for Scientific  
Research (B)  (No.10440073 and No.11695029).

\end{document}